\documentclass{article}

\usepackage{PRIMEarxiv}

\usepackage[utf8]{inputenc} 
\usepackage[T1]{fontenc}    
\usepackage{hyperref}       
\usepackage{url}            
\usepackage{booktabs}       
\usepackage{amsfonts}       
\usepackage{nicefrac}       
\usepackage{microtype}      
\usepackage{lipsum}
\usepackage{fancyhdr}       
\usepackage{graphicx}       
\graphicspath{{media/}}     
\usepackage{tabularx}
\usepackage{caption}
\usepackage{float}

\title{Analysis of market efficiency in main stock markets: using Karman-Filter as an approach
}

\author{
  Beier Liu \\
  Stern School of Business\\ 
  New York University\\ 
  New York \\
  \texttt{bl3619@stern.nyu.edu} \\
   \And
  Haiyun Zhu \\
  Nanyang Business School \\
  Nanyang Technology University\\ 
  Singapore \\
  \texttt{haiyun001@e.ntu.edu.sg} \\
}

\begin{document}
\maketitle

\begin{abstract}
In this study, we utilize the Kalman-Filter analysis to assess market efficiency in major stock markets. The Kalman-Filter operates in two stages, assuming that the data contains a consistent trendline representing the true market value prior to being affected by noise. Unlike traditional methods, it can forecast stock price movements effectively. Our findings reveal significant portfolio returns in emerging markets such as Korea, Vietnam, and Malaysia, as well as positive returns in developed markets like the UK, Europe, Japan, and Hong Kong. This suggests that the Kalman-Filter-based price reversal indicator yields promising results across various market types.
\end{abstract}


\section{Introduction}
\subsection{A Review of Short-term Reversal and Market Efficiency}
Modern financial theory rests on the efficient market hypothesis (EMH), which assumes that investors are rational. Empirically, EMH has been challenged by some ``anomalies'', and with the rise of behavioral finance, the traditional ``market efficiency'' hypothesis has been called into question. Although stock trading should be based on stocks' intrinsic values, many investors act impulsively and detract from the market's efficiency. Noise trading and herding behavior are two examples of common irrational actions. It rejects the ``rational economic man'' assumed in conventional economic and financial research in favor of the more nuanced ``bounded rationality'' assumption and the individualization of investors.

Mid-to long-term stock price overreaction was first noticed and proposed by DeBondt \& Thaler (1985)\cite{DeBondt1985} in the US stock market. They proposed momentum trading and reversal trading, and show that using the latter during an overreaction will lead to a positive stock return. Furthermore, for more than 40 years, the short-term return reversal has been a recognized phenomenon. According to Jegadeesh (1990)\cite{Jegadeesh1990}, a reversal strategy that buys and sells stocks based on their previous month's returns and holds them for a month produced profits of roughly 2\% per month between 1934 and 1987.

Some research attributed the inefficiency of the stock market due to the irrationality of retail investors. Generally speaking, institutional investors have more resources, more expertise, and easier access to transaction data than retail investors do. Due to institutional investors' higher herding behavior and feedback trading tendencies than individual investors, Dennis \& Strickland (2002)\cite{Dennist2002} showed that equities having a bigger fraction of fund holdings in the market tended to have higher returns. And groups of individual investors often create stock price swings, whereas institutional investors help to stabilize prices. Similar results were found in follow-up studies by Barber, Odean, and Zhu (2009)\cite{Barber09}.

\subsection{ A Brief Overview of Market Efficiency of Emerging Markets}
Much less research into EMH has been done in developing markets. Moreover, emerging markets are believed to have higher friction, lower liquidity, and asymmetric information due to less efficient information transmission and less reliable institutional information (Bekaert and Harvey, 2002\cite{Bekaert2002};Griffin, Kelly, and Nardari,2010\cite{Griffin2010}). Because of this information deficiencies in emerging nations, the newly established stock market is inefficient, and market efficiency generally follows the norms of dynamic evolution (Cornelius, 1993\cite{Cornelius1993}).

Recently, only a few of studies have put EMH to the test in isolated emerging economies. However, limitations associated with small sample sizes render these investigations unconvincing. The Indian market was found to be inefficient by Hiremath and Kumari(2014)\cite{Hiremath2014}, although they also found that it was improving with time. For the Czech market, Arltova (2000)\cite{Arlt2000} employed variance ratio tests to disprove the random walk hypothesis. When Seth and Sharma (2015)\cite{Seth2015} examined the informational efficiency and integration of several markets in Asia and the US, they found weak inefficiencies. In an effort to reconcile behavioral finance and market efficiency, Xiong et al. (2019)\cite{Xiong2019} demonstrated that while the Chinese stock market is efficient, the "Adaptive Market Hypothesis" provides a more accurate representation of the market. To date, there has been a lack of specificity in the research on improving market efficiency.

\subsection{The Basic Idea of Testing the Market Efficiency}
According to the efficient market hypothesis (weak-form), stock prices fluctuate randomly or that it is impossible to forecast future prices based on historical data. Tests for independence are used to assess returns in order to assess this hypothesis. A martingale sequence is frequently used in probability theory to characterize a generalized random walk. The variance-ratio methodology (Charles and Darné, 2009\cite{Charles2009}), the autoregressive time series test (test of randomness including the runs test), and the unit root tests (Dickey–Fuller, Phillip Perron, Ng–Perron, Elliot–Rothernbert–Stock) have all been used in previous studies.

However, traditional tests perform very poorly, according to Monte Carlo studies by Elliot, Rothenberg, Stock (ERS), and NG–Perron, among others. Additionally, several studies addressed this by using a different test called BDS in order to properly study independently and identically distributed (IID) data, as traditional tests did not take into account the nonlinearity of the data (Broock et al., 1996\cite{Broock1996}).

\subsection{Why We Choose Kalman Filter as a Test of Market Efficiency}
There are several ongoing debates on how to best measure market efficiency, notwithstanding recent advances in this area. Beginning in the 1980s, academics began exploring the impact of investor behavior on market fluctuations. Shiller (1981)\cite{Shiller1981} noticed that S\&P 500 index's return changed whenever fresh information entered the market. With the phrase ``adaptive market hypothesis'' (AMH), Lo provided a novel explanation for efficiency in financial markets in 2004. According to Lo's theory, the market may take some time to adjust to changing conditions since investor behavior is not viewed as purely rational but rather as the result of a number of interrelated psychological elements (for examples, see Barber and Odean (2000)\cite{Barber2000} and Shleifer (2000)\cite{Shleifer2000}). Studies of investors' attitudes are a modern addition to this field of study.

Following Lo's development of the idea, methods for examining market efficiency underwent modifications. It appears that perceptions of the degree of market efficiency have shifted as a result of advancements in machine learning information extraction algorithms and the accessibility of both quantitative and qualitative data, such as that found on social media. 

As a result, it's possible that the confirmed markets' efficiency has since been verified. As a result, the field of finance has a wealth of literature devoted to modeling and forecasting financial time series using statistics and mathematics. The techniques employed range from traditional statistical and econometric approaches to more contemporary machine learning (ML) and evolutionary-based optimization strategies; see e.g. Kanas and Yannopoulos (2001)\cite{Kanas2001} and Rather et al. (2015)\cite{Rather2015}.

Price fluctuations on a daily basis are viewed as a white-noise process, or Brownian motion, in terms of modern financial engineering principles. Because white noise is a random process in which successive values are independent of each other, it follows that daily price increases and decreases have an equal chance of occurring. In actuality, it is not unusual for a market item to experience a stream of consecutive down days or up days over a brief period of time. The prices are said to be correlated when such event occurs. Our goal is to employ a Kalman Filter on these correlations to foretell future price changes. We will build upon our prior work here by using a Kalman Filter in place of our previous simple one-day predictor. As used here, the Kalman Filter is a two-stage technique predicated on the assumption that the data contains a smooth trendline representing the true value of the market item before being affected by market noise. And this filter be used to forecast stock price movement. A previous paper by the author (Li X. and Li W., 2000)\cite{Li2001} has explored the evolution of China's stock market efficiency for the period between May 1992 and March 2000.

\subsection{What We Have Found in the Data}
In our analysis, we may begin by introducing the data and the fundamental concept of Karman Filter analysis. Next, we may present the results of evaluations of market efficiency on the major stock exchanges. Finally, for developed and emerging stock markets, the short-term and long-term adaptation of market efficiency.

Test on Market Efficiency: After constructing portfolio from Kalman Filter model and checking backtest statistics, it is easy to verify that the Kalman Filter strategy doesn’t work in US but works well in emerging markets such as Korea, Malaysia and Vietnam. Meanwhile, the strategy performance decays significantly in UK, Europe, Japan, Hong Kong in recent years, which may indicate improvement of price discovery efficiency in those markets.

\section{Data and methodology}
We obtain the constituents of each stock exchange's main board index from WIND and the exchange official website. And we then obtain the stock price from Yahoo Finance. Exchanges characteristics are shown in \ref{table:1}.

\begin{center}
\begin{tabularx}{0.9\textwidth} { 
  | >{\raggedright\arraybackslash}X 
  | >{\centering\arraybackslash}X 
  | >{\centering\arraybackslash}X 
  | >{\centering\arraybackslash}X 
  | >{\raggedleft\arraybackslash}X | }
 \hline
 country code & exchange type & main board index & number of stocks & data history \\
 \hline
 US  & DM & SPY  & 504 & 2017.01 - 2024.03  \\
 UK  & DM & FTSE100  & 100 & 2017.01 - 2024.03  \\
 EU  & DM & STOXX500  & 500 & 2017.01 - 2024.03  \\
 JP  & DM & TOPIX2000  & 2159 & 2017.01 - 2024.03  \\
 HK  & DM & HSCI+HSI  & 523 & 2017.01 - 2024.03  \\
 KR  & EM & KOSPI1000  & 1000 & 2017.01 - 2024.03  \\
 VN  & EM & VNX  & 415 & 2023.01 - 2024.03  \\
 MY  & EM & MYX  & 95 & 2017.01 - 2024.03  \\
 ESG  & EM & CB  & 168 & 2017.01 - 2024.03  \\
\hline
\end{tabularx}
\smallskip
\captionof{table}{Key Information of Stocks Exchanges.}
\label{table:1}
\end{center}

\subsection{Dataset}
We use Yahoo finance api to download stock market data (included open, high, low, close, volume) from 2018.01 to 2024.03. We have collected the market data for most comprehensive universe/main board index constituents for major developed markets (US, UK, Europe, Japan, Hong Kong), and emerging markets (Korea, Malaysia and Vietnam).

\subsection{Karman Filter}
The Kalman Filter is a recursive algorithm developed in the 1960s to estimate the future stage of a moving target by measuring its current position. When applied to financial market data, this technology transforms noisy price observations into a sequence of intrinsic value hidden states. This implies that the estimated result and the current observation alone can be used to calculate the estimate for the current state. It does not require any prior estimates or observations, in contrast to batch estimation techniques.

Two variables are used to represent the filter's state: 2) $P_{k|k}$, the a posteriori estimate covariance matrix (a measure of the estimated accuracy of the state estimate); 1) $X_{k|k}$, the a posteriori state estimate mean at time k given observations up to and including at time k. We will refer to the state transition matrix $F_k$, measurement matrix $H_k$, and observation variable $z_k$ for clarity's sake. 

The alpha-beta filter and the Kalman-Filter share a similar algorithmic structure. Despite the fact that the Kalman-zFilter can only be expressed as one equation,the two stages that are typically considered to be involved are "Predict" and "Update." To be more precise, the predict phase creates an estimate of the state at the current timestep using the state estimate from the preceding timestep.

The equation for (a priori) anticipated state estimation is:
$$x_{k|k-1}  =F_{k}x_{k-1|k-1}+B_{k}u_{k}$$

And the equation for (a priori) estimate covariance estimation is:

$$P_{k|k-1}=F_{k}P_{k-1|k-1}F_{k}^{T}+Q_{k}$$

This predicted state estimate is also referred to as the a priori state estimate since it is an estimate of the state at the current timestep but does not include observational data from the current timestep. During the update phase, the state estimate is refined by adding the innovation (the pre-fit residual) to the previous estimate and multiplying it by the optimal Kalman gain. The term "a posteriori state estimate" refers to this improved estimate that was obtained from the current observation.

Where the estimation of Innovation or measurement pre-fit residual is:
$$y_k=z_k-H_kx_{k|k-1} $$ 
Innovation (or pre-fit residual) covariance $S_k$ can be estimated as:
$$S_k=H_kP_{k|k-1}H_k^T+R_k $$

And the optimal Kalman gain $K_{k}$ can be calculated as 
$$K_k = P_{k|k-1}H_k^TS_{k-1}$$. 
In the update phrase, we will update a posteriori state using estimation below:
$$x_{k|k}=x_{k|k-1}  +K_ky_k$$
And the covariance matrix can be updated as:
$$P_{k|k}=(I-K_kH_k) P_{k|k-1}$$
Where the post-fit residual can be measured as: 
$$y_{k|k}=z_k-H_kx_{k|k}$$
The two stages usually happen in turn, with the update taking into account the observation and the prediction pushing the state forward until the following planned observation. In our estimation, the predicted stock price can be obtained from $x_{k|k-1}$ , and we assume that the noise coveriance matrix is identity matrix.

\subsection{Model and Stock Return Forecast}
For each day and each stock, we will feed the latest close price(C) to Karman-Filter model as observations, and we retrieve the true state value of today’s close ($C_{kalman}$). With the assumption that the next day stock price will mean revert to true state value, our stock return forecast for next day will hence be $(C- C_{kalman} )/((C+C_{kalman})*0.5)$.

\subsection{Portfolio Construction}

We will construct an equity long-short portfolio based off our return forecast. Firstly, in the portfolio construction stage, for each country, we will normalized our stock return forecast to guarantee mean zero and unit standard deviation. Then, we will daily re-balance our portfolio to make sure each stocks’ daily dollar position is proportional to normalized stock return forecast. Following these steps, we will run a portfolio based on our Kalman-Filter price reversal strategy, of which we will calculate simulated back-test statistics.

\section{Results}

We apply the Karman-Filter analysis to both developed and emerging stock markets, including the United States, the United Kingdom, Europe, Japan, Hong Kong, South Korea, Vietnam and Malaysia. Overall, the US market is the most efficient compared to other markets. The y-axis, $ROG(bps)$, represents the daily return on gross value in basis points. $T1$ indicates the daily return on next trading day, and $T_{j}$ means the daily return on $T+j$ trading day. Regardless long-term or short-term, we are unable to profit from the U.S. market. The United States, the United Kingdom, and Japan have the lowest returns, indicating that their markets are the most efficient. And Korea's returns are greater in the extended term. Surprisingly, Vietnam's return decreased over the next five trading days, but increased over the extended term. Despite Hong Kong's high stamp duty, its return is not as high as other stock markets, such as Vietnam and Korea.

\begin{figure}[h]
    \centering
    \includegraphics[width=1.\textwidth]{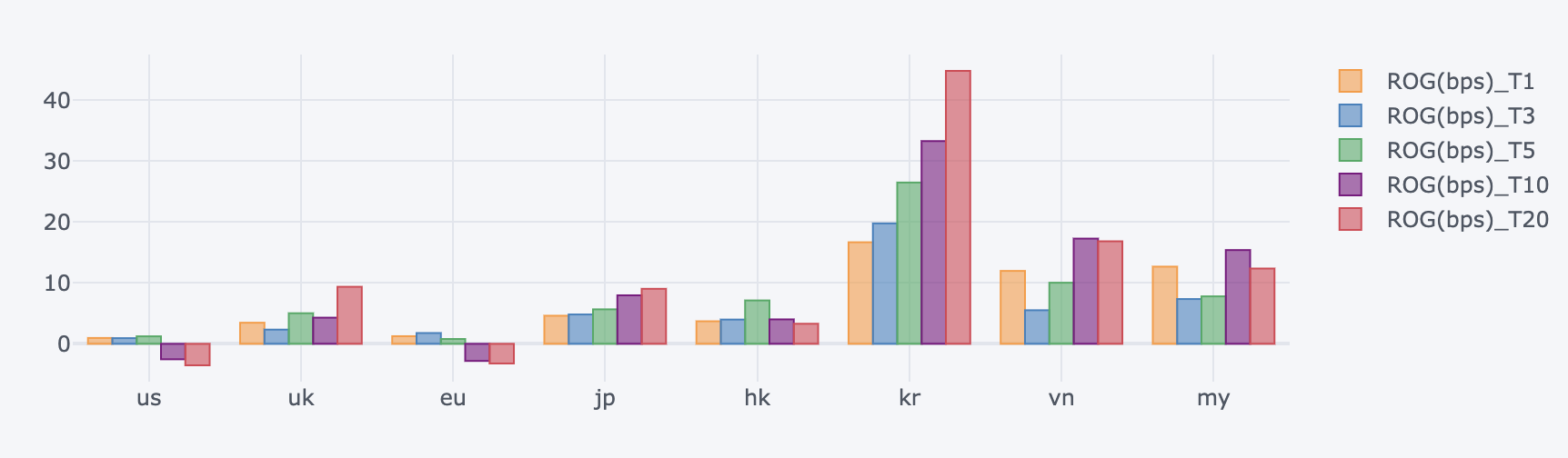}
    \caption{Country Portfolios ROG(return on gross) Plot}
    \label{fig:rog}
\end{figure}

Next, we calculate the stock return for each less efficient stock market to determine if there is a time trend in their daily return, where $execution1$ represents the return from the market open to T+1's market open, $execution2$ represents the return from the market close to T+1's market close, and $execution3$ represents the return from the market open to same day's market close. 
Generally $execution3$ generates the lowest return among all, and $execution1$ generates the highest return. We use metrics like return on gross (ROG) and sharpe-ratio (SR) to compare strategy historical performance.

Since the indication is the measurement of mean reversal to real value anticipated by Karman-Filter on noise price movements, from the images below. Several patterns are discernible.

The three execution strategies all generate positive returns in all markets except for the US consistently after ignoring the returns from the COVID period, which shows that most markets are inefficient for the majority of the time. The first universal ``reversal collapse'' occurs in 2020.03, which is the start of COVID and most international long short hedge funds had major draw downs in the same period.

Additionally, by observing each market separately, we may compare the effectiveness of the three strategies depending on their various execution techniques. $execution1$ approach is typically the most most difficult to execute in real financial markets because, during this open market period, the market is digesting a lot of overnight information in the opening auction, and the market capacity is the smallest when auction trading volume is taken into account. However, due to the lack of intra-day minute-level data, the $execution2$ approximation is a practical approximation of the real market because it accounts for 10\% to 20\% of the daily trading volume and most importantly, it takes all day to execute. In reality, traders frequently use time-weighted average price (TWAP) or volume average price (VWAP) and trade in a certain interval to eliminate the price impact. And the $execution3$ return can be viewed as the difference between the returns of the $execution1$ and a hyper-theoretical overnight portfolio. Therefore, the ability to arbitrage should decline first with $execution2$ and then $execution1$ (which is ordered by the market capacity) if the trading signal of this Kalman Filter strategy decays, which indicates that the market is becoming more efficient.

\begin{figure}[h]
    \centering
    \includegraphics[width=1.\textwidth]{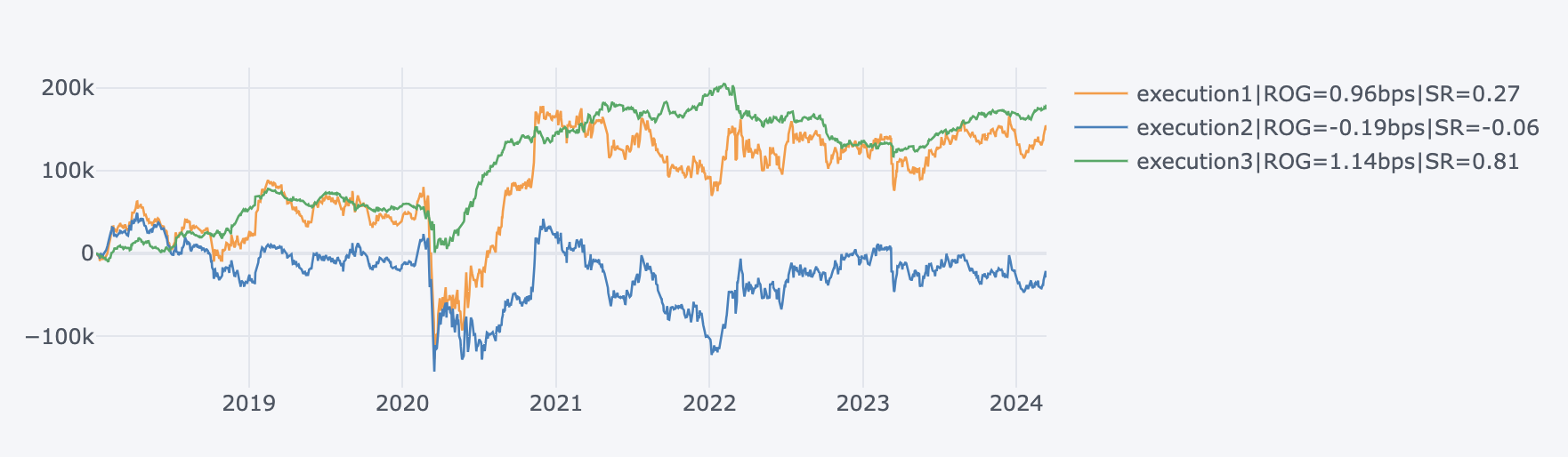}
    \caption{Strategy Backtest in USA}
    \label{fig:us}
\end{figure}

The various patterns in the worldwide markets shown below show that the US market \ref{fig:us} (ROG -0.12bps, SR -0.04) is the most effective because even in $execution1$ strategy, where the market capacity is the smallest, there is a tiny amount of room for arbitrage. 

\begin{figure}[h]
    \centering
    \includegraphics[width=1.\textwidth]{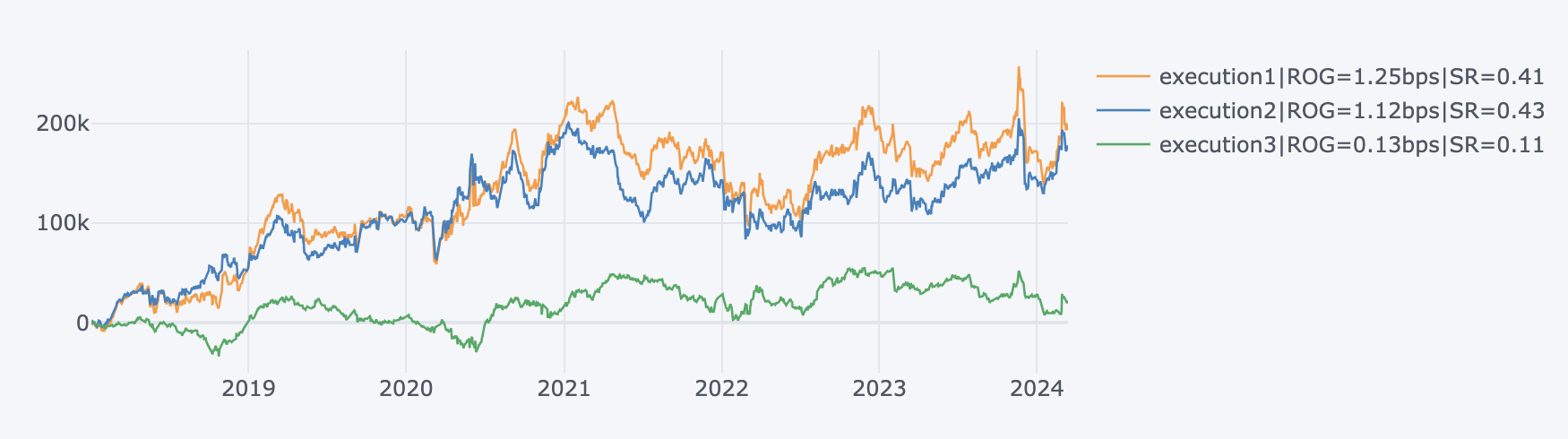}
    \caption{Strategy Backtest in Europe}
    \label{fig:eu}
\end{figure}

\begin{figure}[h]
    \centering
    \includegraphics[width=1.\textwidth]{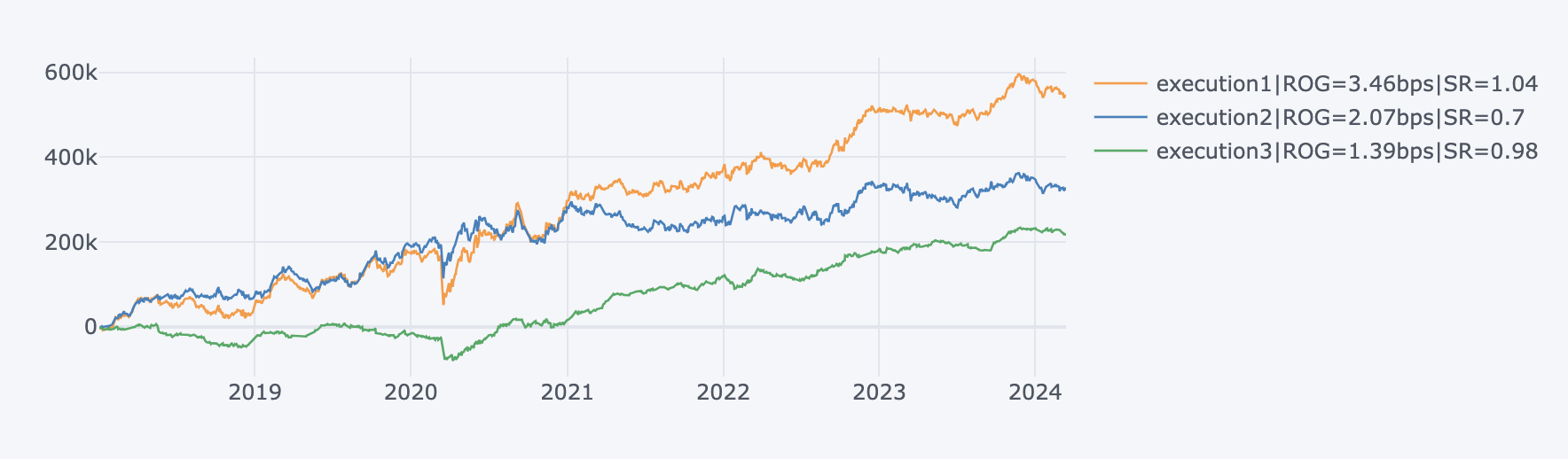}
    \caption{Strategy Backtest in United Kingdom}
    \label{fig:uk}
\end{figure}

\begin{figure}[h]
    \centering
    \includegraphics[width=1.\textwidth]{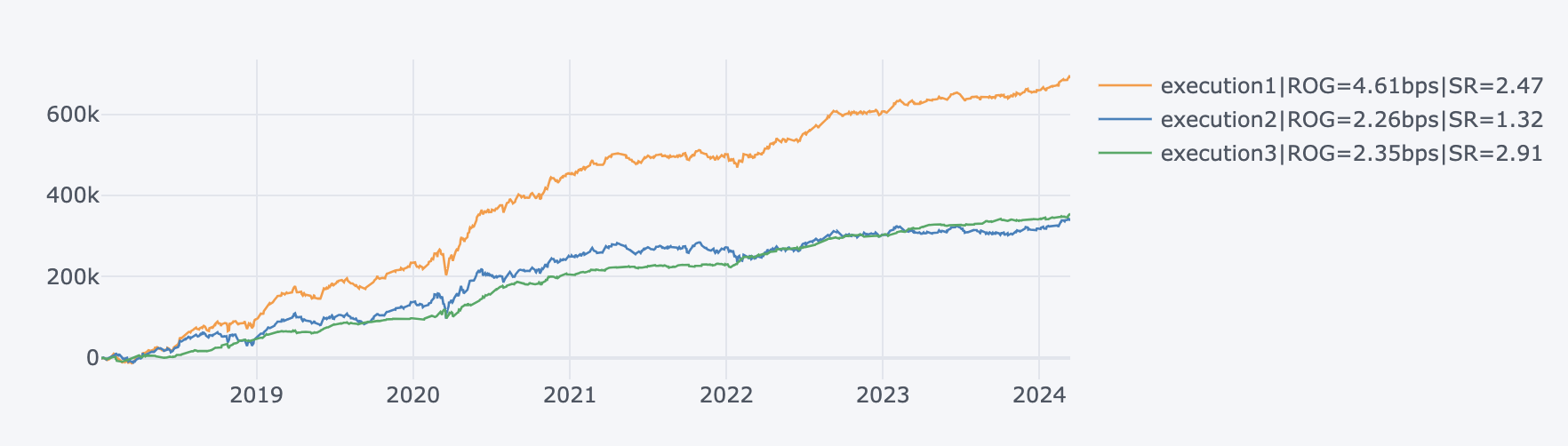}
    \caption{Strategy Backtest in Japan}
    \label{fig:jp}
\end{figure}

\begin{figure}[h]
    \centering
    \includegraphics[width=1.\textwidth]{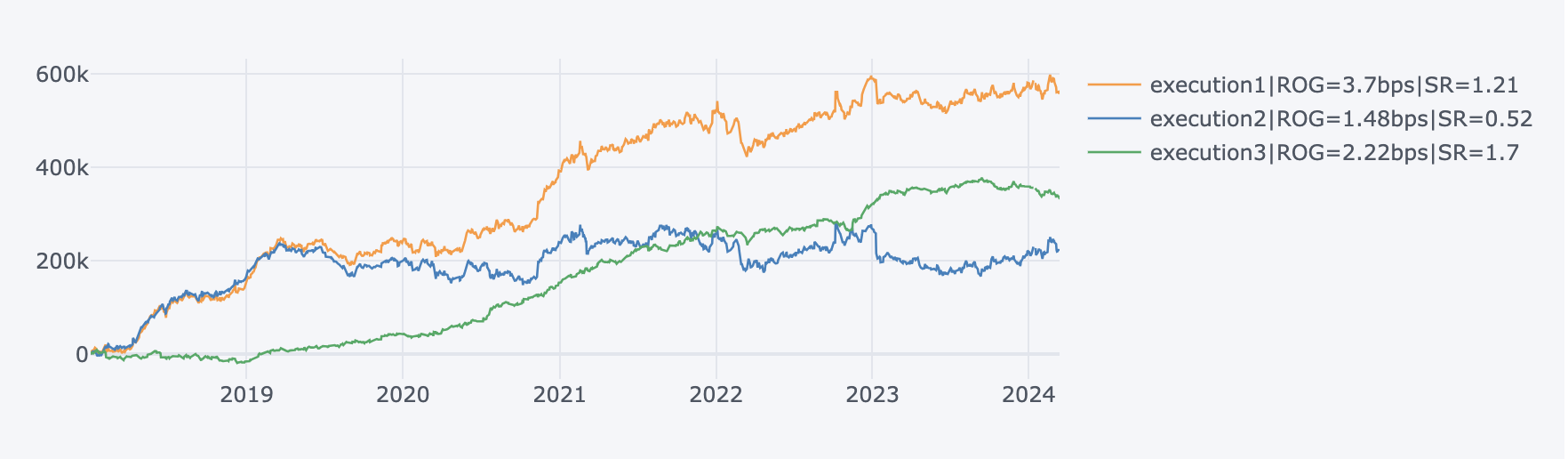}
    \caption{Strategy Backtest in Hong Kong}
    \label{fig:hk}
\end{figure}

\begin{figure}[H]
    \centering
    \includegraphics[width=1.\textwidth]{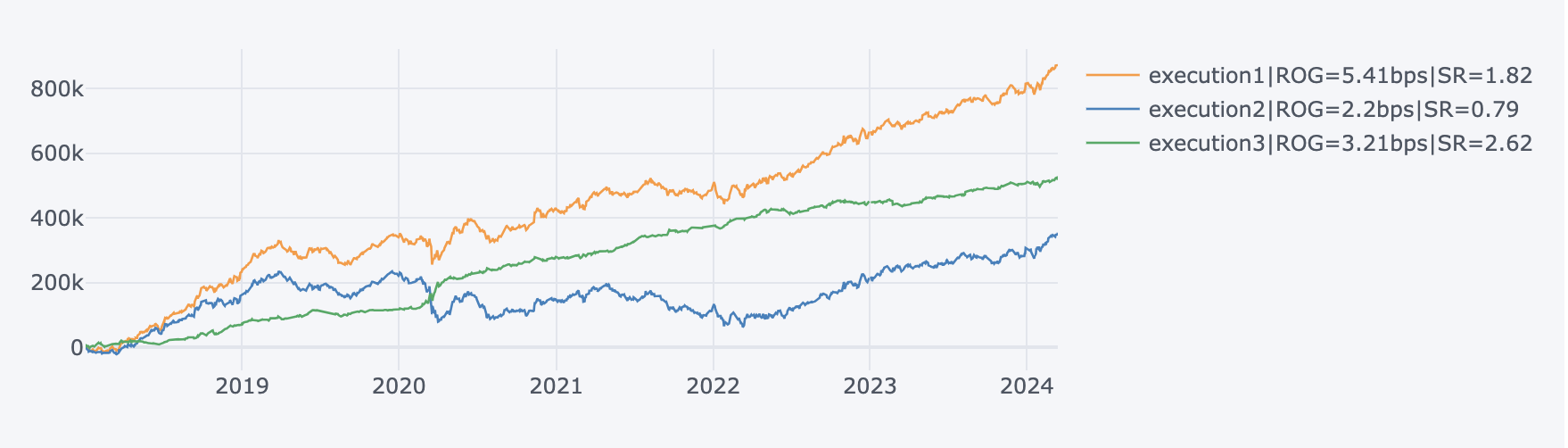}
    \caption{Strategy Backtest in ESG Index}
    \label{fig:esg}
\end{figure}

And for Europe(ROG 1.15bps, SR 0.47) Fig\ref{fig:eu}, the United Kingdom(ROG 2.25bps, SR 0.75) Fig\ref{fig:uk}, Japan(ROG 2.32bps, SR 1.32) Fig\ref{fig:jp}, Hong Kong(ROG 1.29 SR 0.45) Fig\ref{fig:hk}, and ESG indexes(ROG 1.92, sr 0.69) Fig\ref{fig:esg}, we can see that the $execution1$ strategy hardly ever makes a profit while the $execution2$ strategy has a limited amount of room for arbitrage when the market capacity is smaller. And out of the four markets, Hong Kong and Japan are the most alluring due to the largest divergence between the $execution1$ and $execution2$ strategies, specifically the higher return of the $execution3$ strategy (green line), and the highest SR ratio among the four markets.


\begin{figure}[H]
    \centering
    \includegraphics[width=1.\textwidth]{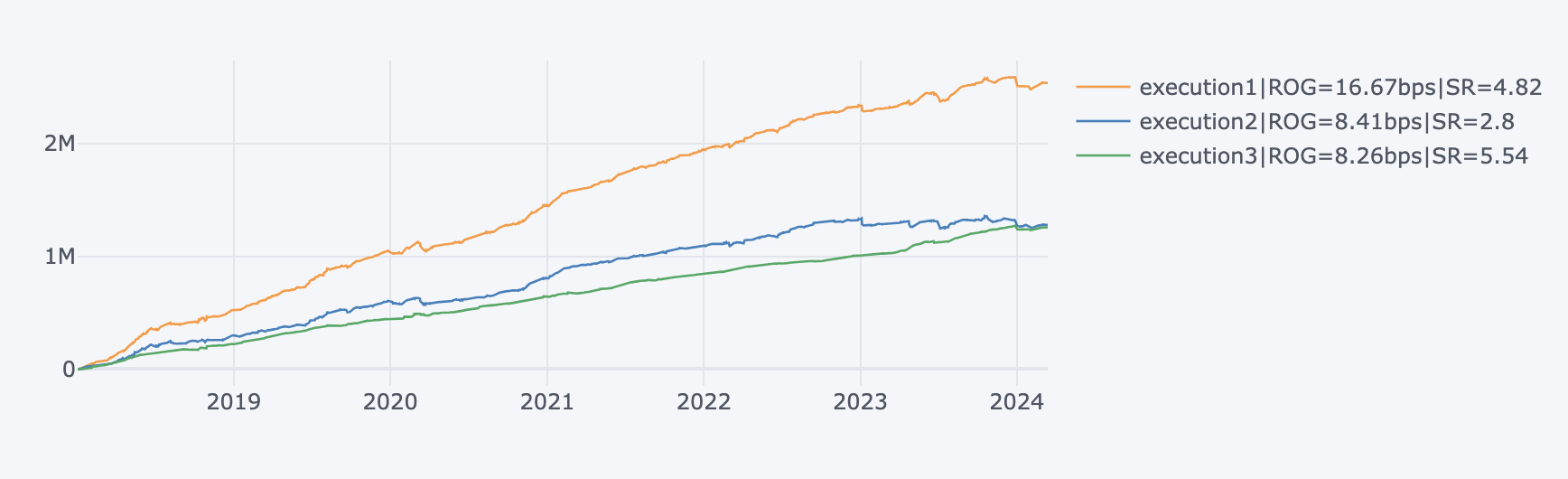}
    \caption{Strategy Backtest in Korea}
    \label{fig:kr}
\end{figure}

\begin{figure}[H]
    \centering
    \includegraphics[width=1.\textwidth]{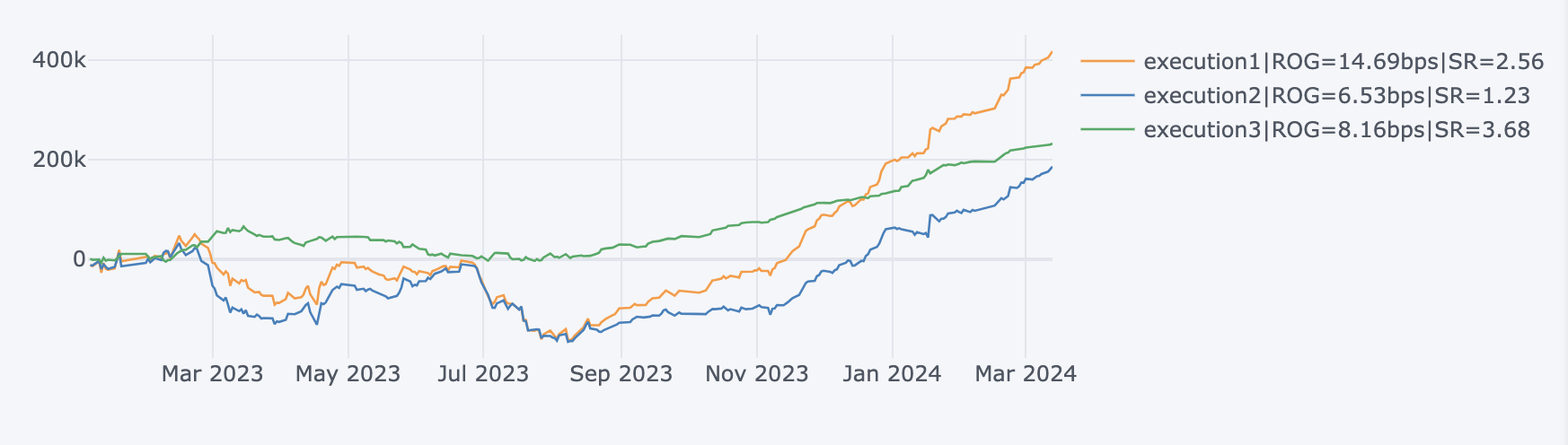}
    \caption{Strategy Backtest in Vietnam}
    \label{fig:vn}
\end{figure}

\begin{figure}[H]
    \centering
    \includegraphics[width=1.\textwidth]{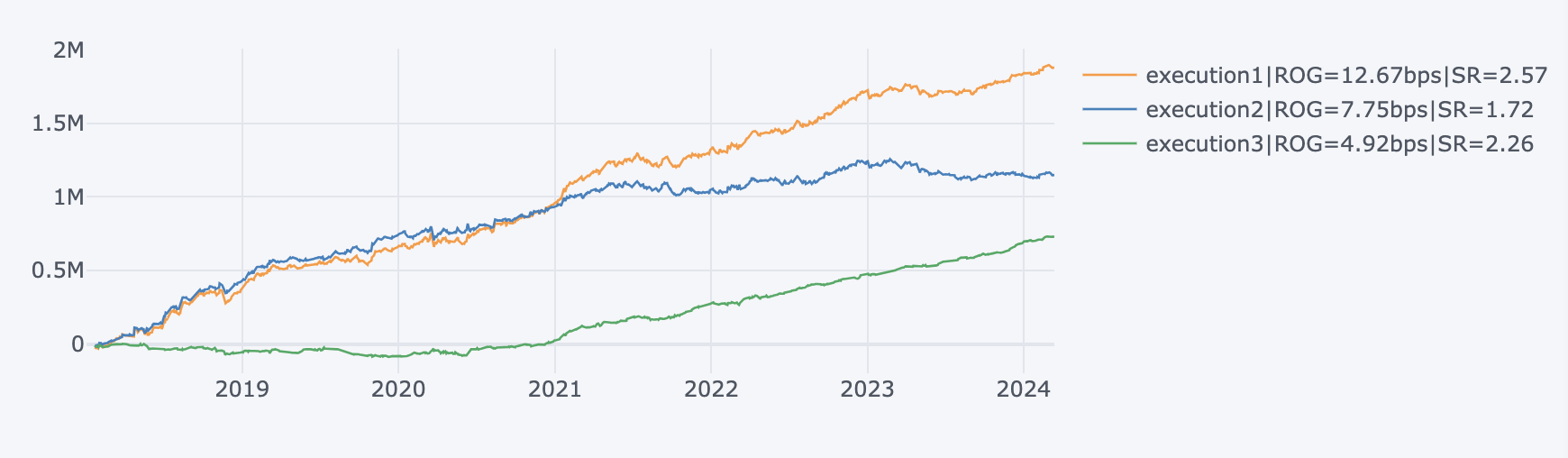}
    \caption{Strategy Backtest in Malaysia}
    \label{fig:my}
\end{figure}

Aside from that, the least efficient markets include those in Korea (ROG 9.33, SR 3.18) Fig\ref{fig:kr}, Vietnam(ROG 19.01, SR 2.73) Fig\ref{fig:vn}, Malaysia(ROG 12.76, SR 2.51) Fig\ref{fig:my} since there is still room for arbitrage in open auction, regular trading, and intraday trading. Overall, from the standpoint of efficiency, we can see that the market is becoming more efficient, even in developed countries.


\section{Conclusion and Discussion}

In this paper, we applied the Kalman-Filter analysis to test the market efficiency of major stock markets across developed and emerging markets. The key findings are:

First, even in execution techniques with smaller market capacity, the Kalman-Filter approach does not yield appreciable gains, suggesting that the U.S. market is the most efficient. Although there is some degree of inefficiency in other established markets, such as the UK, Europe, Japan, and Hong Kong, the execution2 strategy—which is based on closing the market close to the next day's close—generates statistically positive returns albeit tiny ones. With greater return divergences across various execution techniques, Hong Kong and Japan stand out. All three execution strategies show significant return premiums for emerging countries, such as Korea, Vietnam, and Malaysia, suggesting a higher degree of market inefficiency in these areas.

Moreover, when returns of the Kalman-Filter approach is decreasing, particularly during times like the COVID-19 outbreak in 2020, there are indications that market efficiency is improving over time, even in the inefficient markets.

Overall, the Kalman-Filter method looks at how asset prices mean-revert from their fundamental values, offering a fresh perspective on how to measure market efficiency overtime. And our findings suggest that while market efficiency varies across countries, emerging markets tend to be less efficient than developed ones.

\appendix

\bibliographystyle{elsarticle-num} 
\bibliography{references}






\end{document}